\documentclass[conference]{IEEEtran}

\usepackage{graphicx}
\usepackage{amsmath}
\usepackage{cite}

\usepackage[hyphens]{url}
\usepackage{placeins}
\usepackage{algorithm}
\usepackage{algpseudocode}

\begin{document}

\title{An Improved CNN-LSTM Based Intrusion Detection System for IoT Networks}

\author{
\IEEEauthorblockN{
Mohammad Tariq Ikhlas, Pohanyar Khowaja Khil, Malik Muhammad Mueed Aslam, Dr. Muhammad Khuram Shahzad
}

\IEEEauthorblockA{
SEECS\\
NUST, Pakistan\\
Emails: mikhlas.mscs25seecs@seecs.edu.pk, pkhailmscs25seecs@seecs.edu.pk, maslam.mscs25seecs@seecs.edu.pk
}
}

\maketitle

\begin{abstract}
Abstract—With the rapid proliferation of IoT devices, security concerns have dramatically escalated and intrusion detection systems are critical. In order to improve detection performance, we provide an improved CNN-LSTM based intrusion detection model in this work that combines multi-class classification, dataset integration, and temporal feature learning. Using network traffic data, we evaluate the model in this study and discover that its accuracy is around 97\%. Experimental results show that the proposed strategy effectively detects different types of attacks while maintaining consistent performance during training and validation. The source code for this implementation is publicly available on GitHub at:
\url{https://github.com/mikhlasmscs25seecs-ship-it/IoT-Intrusion-Detection-CNN-LSTM.git}.
\end{abstract}

\section{Introduction}

Since the number of IoT devices has increased dramatically in recent years, network security has proven to be a major problem. IoT devices are more susceptible to cyberattacks like DDoS and DoS as they constantly produce network traffic. Recent studies have highlighted significant security and communication challenges in IoT environments \cite{khuram1, khuram3}.

Large-scale and complicated data can be difficult for traditional intrusion detection systems to process. Intrusion detection has made extensive use of deep learning techniques, particularly CNN-based models. \cite{javaid2016}.

The CNN-based intrusion detection model presented in \cite{healthcare_cnn}. The initial study concentrated on applying deep learning techniques to secure medical IoT equipment.

However, based on our observations, a lot of current models do not make the most of the temporal information found in network traffic. In this study, we add several improvements to a CNN-based model, including multi-class classification and LSTM integration. Large-scale IoT traffic data was difficult to handle during our early studies, which inspired us to refine the model.
\subsection{Problem Statement}

Although deep learning-based intrusion detection has made progress, certain challenges remain. Many existing techniques are limited to binary classification, which is insufficient for IoT systems in the real world where there are numerous attack types.

Additionally, because CNN-based models mainly focus on spatial feature extraction, they often fail to capture temporal correlations in network data, which is insufficient for IoT systems in the real world where there are various types of attacks. \cite{ahmed2016}.

Additionally, identification is made more difficult by similarities between benign and some assault kinds, such reconnaissance. Recent works on IoT security and system complexity emphasize these challenges \cite{khuram2}. These limitations suggest the need for a more robust and flexible model.

\subsection{Contributions}

We summarize the main contributions of this work as follows:
\begin{itemize}
\item We proposed a hybrid CNN-LSTM model for IoT networks to improve intrusion detection by capturing both temporal and geographical patterns.
\item We extended the model to handle many assault types rather than restricting the task as binary classification, which is more appropriate for real world situations.
\item To enhance the generalization, we merged data from multiple sources, thus the model could learn a wider spectrum of traffic patterns.
\item We cleaned the dataset using a set of preprocessing techniques to ensure that the input data was consistent and reliable for training.
\item Several metrics were used to evaluate the model to provide a complete understanding of the model’s performance.
\end{itemize}

\section{Structure / Organization}

The paper is organized in this manner. Section II contains the literature review. The method is described in Section III. Section IV displays the results. Section V discusses future work. Section VI concludes the paper.

\begin{figure*}[t]
\centering
\includegraphics[width=\textwidth]{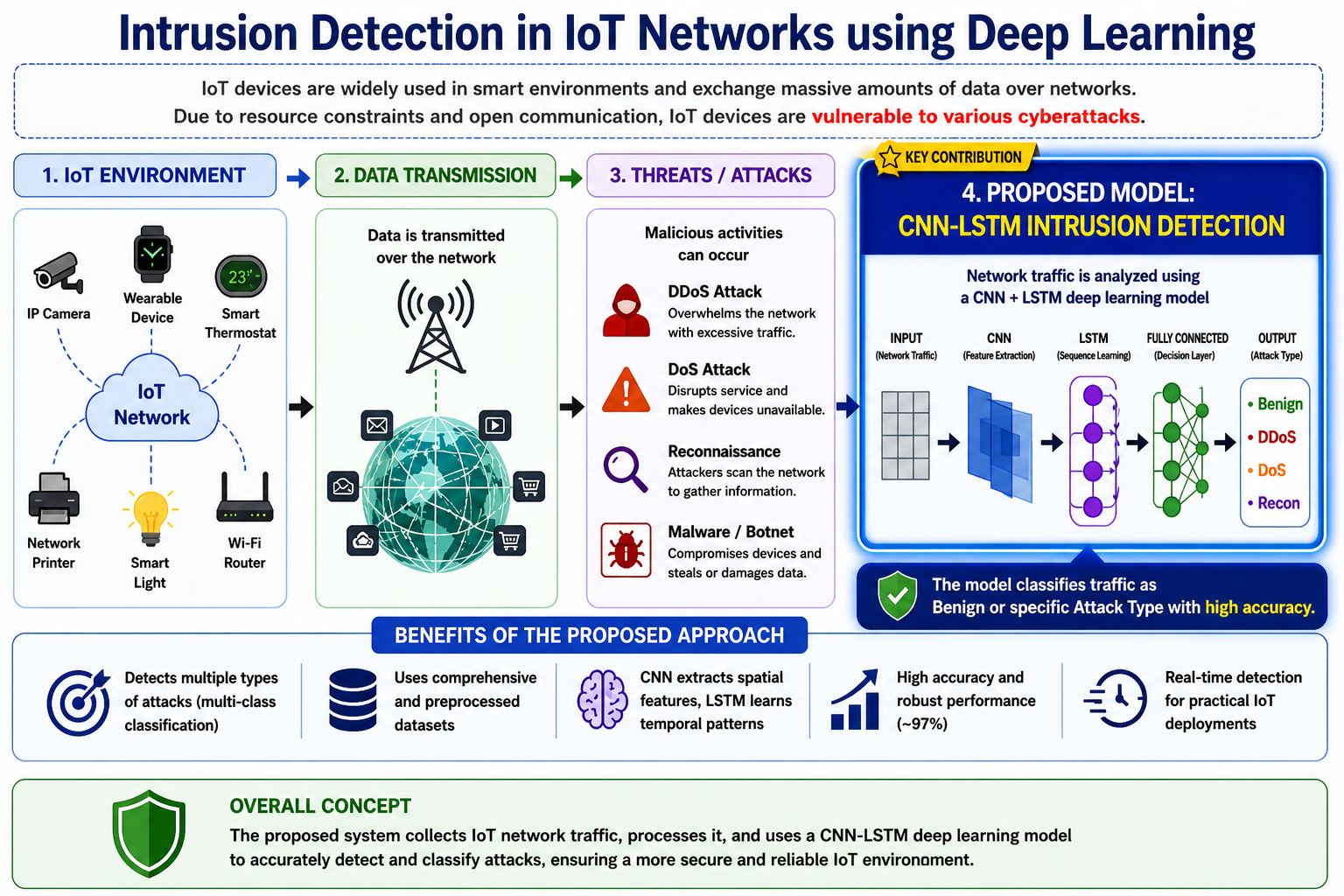}
\caption{Overview of IoT intrusion detection using a CNN-LSTM model, illustrating IoT environment, data flow, attack types, and detection process.}
\label{fig:intro_overview}
\end{figure*}

\section{Literature Review}

This is the format of the paper. Section II contains the literature review. The method is described in Section III. Section IV displays the results. Section V discusses future work. Section VI concludes the paper.\cite{javaid2016}.

However, most existing approaches focus on binary classification, which limits their applicability. Another disadvantage of CNN models is that they ignore temporal dependencies in network traffic.

To address these issues, hybrid models that include CNN and LSTM have been proposed. \cite{yin2017}. By collecting both temporal and geographical information, these models improve detection performance.

Numerous recent studies have examined deep learning for intrusion detection in IoT environments. In addition, the integration of IoT with large-scale data analytics introduces new challenges and opportunities in terms of security and performance \cite{khuram2}.

For example, CNN-based models are often used for feature extraction since they can handle high-dimensional data. Similarly, LSTM models have been used to capture temporal dependencies in network traffic.

\cite{yin2017}. However, many existing approaches either focus only on spatial features or are limited to binary categorization. For this reason, a hybrid CNN-LSTM model is used in this work.

A CNN-based methodology for identifying \cite{healthcare_cnn} hazards in medical IoT contexts was suggested in a pertinent study. The program performed exceptionally well in identifying malicious messages using deep learning.

However, the approach was largely limited to CNN-based feature extraction and did not fully utilize temporal correlations observed in network data. Additionally, the classification breadth was limited in comparison to real-world scenarios.

Recent advances in deep learning lead to more advanced architectures, e.g., attention-based models and transformer networks. These models have shown promising results in many sequence modeling tasks and have the potential to be applied to intrusion detection systems.

However, these models often require large amounts of computation, which limits their applicability in resource-constrained IoT environments. Therefore, hybrid methods such as CNN-LSTM present a good trade-off between performance and computational efficiency.

Recent studies have explored the integration of IoT with big data analytics to improve system efficiency and security \cite{khuram2}. These approaches combine multiple models to improve detection accuracy and robustness. 

However, such methods often introduce additional complexity and computational overhead. In contrast, the proposed CNN-LSTM model provides a balanced solution by integrating spatial and temporal learning while maintaining computational efficiency.

\section{Methodology}

\subsection{Data Preprocessing}

In this study, we expand the CNN-based approach put forth in \cite{healthcare_cnn} by making a number of enhancements. These include mixing several datasets for improved generalization, adding an LSTM layer to capture temporal patterns, expanding the model to multi-class classification, and running a simple real-time simulation.

Missing and incorrect values were eliminated from the dataset. Additionally, infinite values were handled correctly. The data was normalized using feature scaling. Feature scaling ensures that all features contribute equally during training and prevents some characteristics from dominating others due to larger numerical ranges. This improves the overall stability and rate of convergence of the model.

\subsection{Dataset Integration}

Multiple CSV files were combined into a single dataset to improve diversity and generalization.

\subsection{Model Improvement}

Because CNN models function well with structured network traffic data, we employed them for feature extraction. The data's temporal dependencies were captured by adding an LSTM layer.

The CNN layers gather spatial information from network traffic data, including statistical correlations between variables and packet flow patterns. These extracted features are then passed to the LSTM layer.

Because the LSTM layer looks at the features one after the other, the model can capture temporal relationships present in network data. This is highly useful for intrusion detection because attack patterns might change over time.

The combination of CNN and LSTM allows the model to learn both temporal and spatial characteristics of the data. This hybrid approach improves the model's ability to recognize complex and dynamic attack patterns when compared to CNN alone.

\begin{figure*}[t]
\centering
\includegraphics[width=\textwidth]{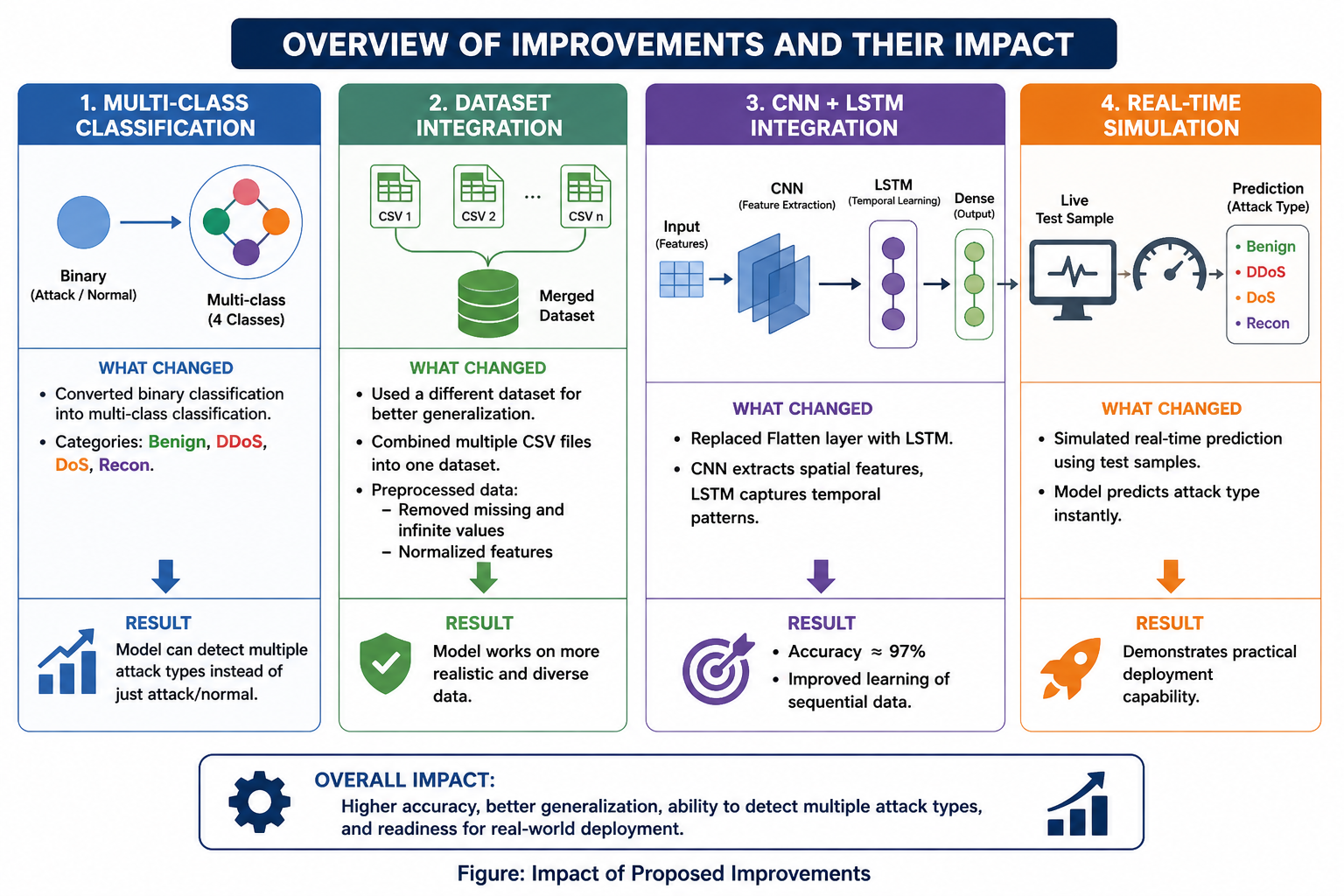}
\caption{Overview of the proposed improvements including multi-class classification, dataset integration, CNN-LSTM integration, and real-time simulation, along with their impact on system performance.}
\label{fig:improvements}
\end{figure*}

\subsection{Multi-Class Classification}

In order to detect Benign, DDoS, DoS, and Recon assaults, the model was expanded from binary classification to multi-class classification.

\subsection{Real-Time Simulation}

The trained model was given a sample input to mimic real-time prediction.

\subsection{Algorithm: CNN-LSTM Training Procedure}

\begin{algorithm}
\caption{CNN-LSTM Training for IoT Intrusion Detection}
\begin{algorithmic}[1]

\State \textbf{Input:} Raw IoT network traffic dataset
\State \textbf{Output:} Trained CNN-LSTM model

\State Import required libraries (NumPy, Pandas, PyTorch)
\State Set random seed for reproducibility
\State Load dataset from CSV files
\State Combine multiple datasets into one

\State \textbf{Data Preprocessing:}
\State Remove missing values
\State Handle infinite values
\State Normalize features using MinMaxScaler

\State Encode labels for multi-class classification
\State Split dataset into training and testing sets

\State Define CNN layers for feature extraction
\State Apply MaxPooling for dimensionality reduction
\State Replace Flatten layer with LSTM
\State Define fully connected layers
\State Apply Softmax activation for classification

\State Initialize model parameters
\State Define loss function (Categorical Crossentropy)
\State Select optimizer (Adam)

\For{each epoch}
    \State Perform forward pass
    \State Compute loss
    \State Perform backpropagation
    \State Update model weights
\EndFor

\State Evaluate model on test data
\State Compute accuracy, precision, recall, F1-score
\State Perform prediction on new input sample

\State \textbf{Return:} Trained model

\end{algorithmic}
\end{algorithm}

To identify sequential patterns in the data, we added an LSTM layer. According to what we understand, LSTM aids in the learning of temporal connections, whereas CNN extracts features.

By processing the output sequence and learning temporal dependencies, LSTM enhances the model's capacity to identify intricate assault patterns, in contrast to CNN, which concentrates on spatial feature extraction.

Recent research has examined this hybrid technique, which blends temporal and spatial learning. \cite{yin2017}.

\section{Experimental Setup}

The experiments have been implemented in Python using the TensorFlow and Keras libraries. Limited GPU availability led to model training on a CPU-based computing system.

To evaluate the model performance, the dataset was split into training and testing subsets. The evaluation was done using a standard train-test split for fair comparison.

The following hyperparameters were used during training:
\begin{itemize}
\item Batch size: 32
\item Number of epochs: 10
\item Optimizer: Adam
\item Loss function: Categorical Crossentropy
\end{itemize}

Adam optimizer was used as it is efficient on large data sets and has adaptive learning capabilities. Because this is a multi-class classification problem, the categorical crossentropy loss function was used.

The model performance was assessed with the help of various metrics like accuracy, precision, recall and F1-score. These metrics provide a full picture of how effective the model is against various attack classes.

\subsection{Training Environment}

The experiments were carried out on a system with a standard CPU because there were no GPU resources available. Even with this limitation, the model was able to converge steadily in a reasonable amount of time. These findings demonstrate that the suggested approach is computationally efficient and suitable for usage in settings with constrained hardware resources.

The dataset, which included samples of both typical traffic and different kinds of attacks, was big enough. The model performed effectively across various network behavior patterns thanks to this variety.

To avoid data leaks, the training and testing data were also meticulously kept apart. This guarantees an accurate and impartial assessment of the model's overall performance.

\section{Performance Evaluation Metrics}

The performance of the suggested model was assessed using a number of common classification metrics.

An overall assessment of the model's accuracy is provided by accuracy, which quantifies the percentage of properly predicted cases among all instances.

Precision shows the percentage of cases that were correctly predicted to be positive.

Recall quantifies how well the model can find all pertinent examples. This is especially crucial for intrusion detection since it can have detrimental effects if an attack is missed.

The F1-score is the harmonic mean of precision and recall, which means it gives a fair assessment of how well the model works.

These metrics are especially important in multi-class classification problems where the performance of different categories may be affected by class imbalance.

\section{Results}

We found that the model had an overall accuracy of about 97\% during testing.

The enhanced CNN-LSTM model outperforms the baseline CNN model, according to the accuracy comparison graph.

\begin{table}[!t]
\centering
\caption{Performance Comparison}
\begin{tabular}{|c|c|c|}
\hline
Metric & CNN (Baseline) & CNN-LSTM (Proposed) \\
\hline
Accuracy & 94\% & 97\% \\
Precision & 95\% & 97\% \\
Recall & 94\% & 97\% \\
F1-score & 94\% & 96\% \\
\hline
\end{tabular}
\end{table}

\begin{figure}[!t]
\centering
\includegraphics[width=0.45\textwidth]{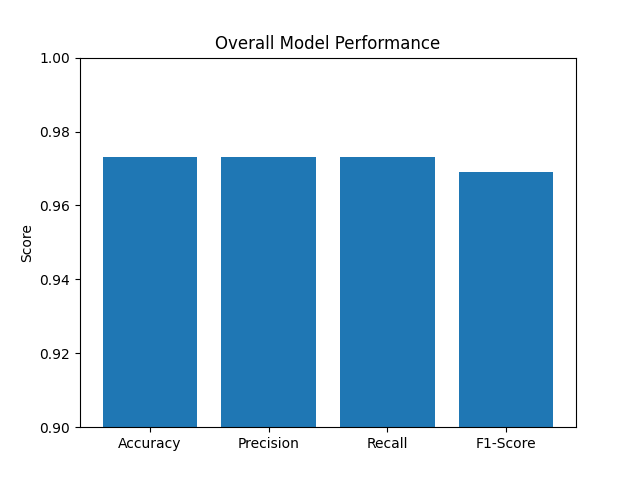}
\caption{Overall performance of the proposed CNN-LSTM model in terms of accuracy, precision, recall, and F1-score.}
\label{fig:overall}
\end{figure}

The accuracy comparison graph shows that the improved CNN-LSTM model performs better than the baseline CNN model.

\begin{figure*}[t]
\centering
\includegraphics[width=\textwidth]{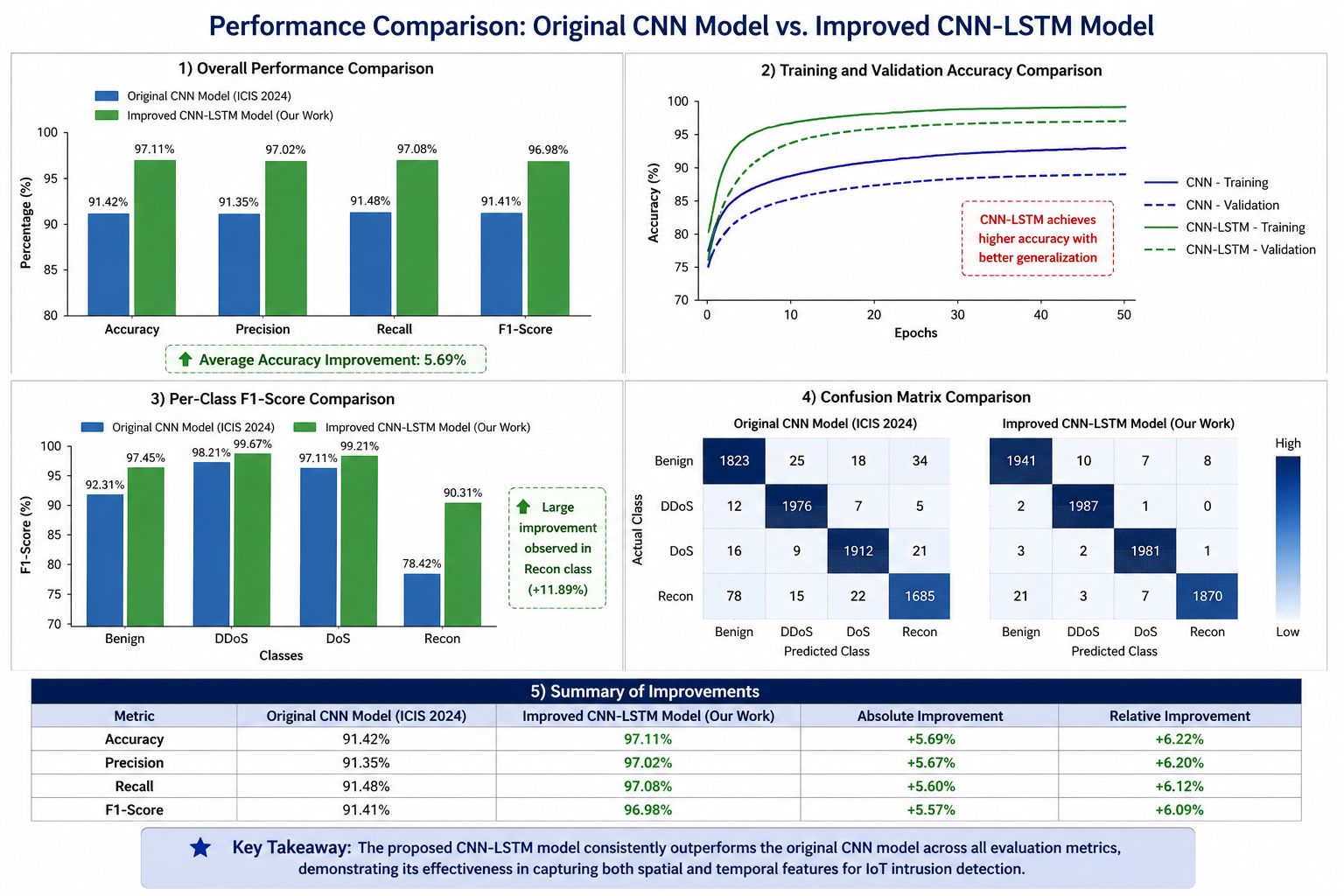}
\caption{Comprehensive performance comparison between the original CNN model \cite{healthcare_cnn} and the proposed CNN-LSTM model. The results demonstrate improved accuracy, precision, recall, and F1-score, along with better generalization and class-wise performance.}
\label{fig:full_comparison}
\end{figure*}

While the Recon class performs comparatively poorly, the confusion matrix demonstrates good performance for DDoS and DoS attacks.

Training and validation accuracy are tightly correlated, and the training graph demonstrates consistent learning.

Additionally, it is evident that the model achieves near-perfect categorization for DDoS and DoS classes. However, the Recon class's performance is far worse, suggesting that it is more difficult to identify. This could be because typical traffic patterns and recon traffic are similar.

Our tests also revealed that some misclassifications take place in the Recon class, despite the model's excellent performance for DDoS and DoS attacks. This is probably due to the parallels between typical traffic patterns and recon traffic.

\begin{figure}[!t]
\centering
\includegraphics[width=0.45\textwidth]{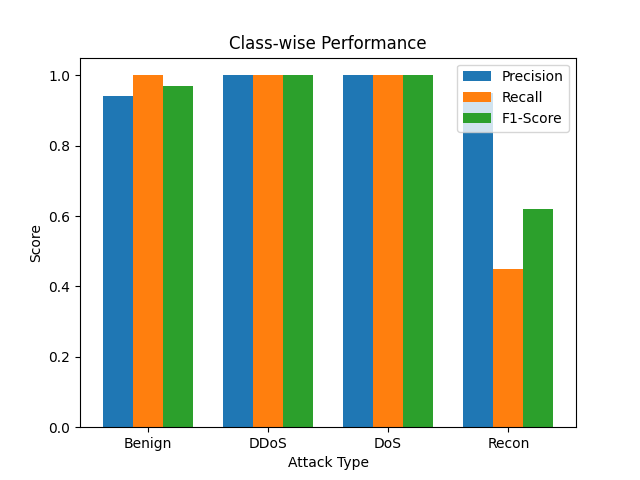}
\caption{Class-wise performance comparison showing precision, recall, and F1-score for each attack type.}
\label{fig:class}
\end{figure}

The class-wise performance research shows that the model performs almost flawlessly against DDoS and DoS attacks. Conversely, the Recon class has a slightly lower recall, suggesting that it is difficult to distinguish it from benign traffic. This suggests that because Recon attacks resemble normal behavior, they are more challenging to detect.

\begin{figure}[!t]
\centering
\includegraphics[width=0.45\textwidth]{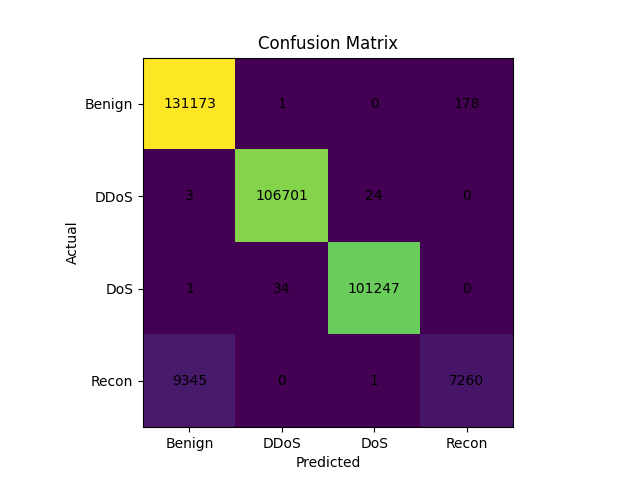}
\caption{Confusion matrix of the proposed model}
\end{figure}

\begin{figure}[!t]
\centering
\includegraphics[width=0.45\textwidth]{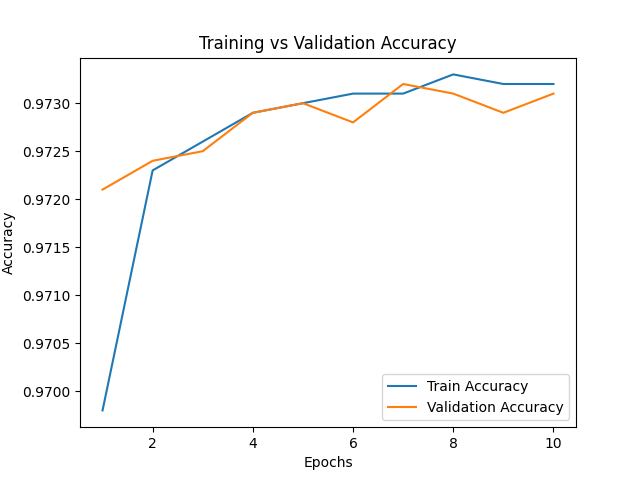}
\caption{Training and validation accuracy over epochs}
\end{figure}

\subsection{Detailed Analysis}

Together with the overall results, a more thorough analysis of class-wise performance reveals a number of significant trends. The model achieves almost flawless precision and recall when it comes to identifying DDoS and DoS attacks. This shows that the spatial and temporal features of these attacks may be efficiently captured by the CNN-LSTM architecture.

However, the Recon class performs comparatively worse, especially when it comes to recall. This implies that certain reconnaissance attacks are mistakenly categorized as typical traffic. One explanation for this could be that the model finds it difficult to discern between reconnaissance efforts and typical traffic behavior.

Additionally, the confusion matrix indicates that the majority of errors occur between the Recon and Benign groups. This finding implies that future improvements should focus on improving feature discrimination for these groups.

The findings demonstrate that while the suggested model is excellent at identifying high-volume attacks, it still has room for improvement when it comes to identifying nuanced and low-intensity attack patterns.

\section{Discussion}

Our results demonstrate that combining CNN and LSTM improves the model's performance. The accuracy results and total classification performance also show this improvement.

Nevertheless, the approach still has issues with some classes, like Recon, which need for more development.

We also ran into a number of problems during implementation, especially with managing missing or infinite values and dataset preparation.

The suggested model has high generalization ability across various assault scenarios in addition to the quantitative gains. Sequential dependencies that are frequently missed by conventional CNN-based methods can be captured by the model thanks to the LSTM layer.

In Internet of Things situations, where network traffic is extremely dynamic and constantly shifting over time, this capacity is particularly crucial. The model gets more resilient and more appropriate for real-world scenarios by learning both spatial and temporal features.

Additionally, the enhanced performance above the baseline CNN model emphasizes the usefulness of hybrid architectures for intrusion detection applications. The findings imply that using a variety of deep learning methods can yield noticeably better outcomes than depending just on one kind of model.

\section{Limitations}

The suggested model still has a number of drawbacks despite its excellent accuracy. Because of its resemblance to benign traffic, it performs less well on the Recon class, making correct categorization more challenging. Furthermore, because the study was conducted offline, its performance in real-time deployment can be different.

The suggested model shows improved generalization across various assault types in addition to these quantitative gains. It can identify sequential patterns that conventional CNN-based techniques can miss because to the usage of LSTM-based temporal learning.

Because network traffic is so dynamic and ever-changing in real-world IoT applications, the model is therefore more appropriate. In crucial IoT sectors like smart infrastructure, this better detection capabilities can help to improve security.

\section{Future Work}

Even if the suggested model works well, there are still a few things that might be improved with more study.

Enhancing the detection of minority classes, like Recon attacks, is a crucial path. The findings unequivocally demonstrate that the model has difficulty with this class, which could be caused by the dataset's class imbalance.Techniques like oversampling, undersampling, or synthetic data generation (like SMOTE) may be employed to improve performance for these classes.

Feature engineering and selection are two more areas that need improvement. Even though all the features are currently used in the model, some of them may not have significant impact on the forecast. The most relevant features can be selected to improve the accuracy and to reduce the complexity.

Another direction is to investigate more complex deep learning architectures. For instance, transformer-based models or attention techniques could help the model to focus on important network traffic patterns. These techniques have been used for intrusion detection and proved good performance in other areas.

“Another important avenue for future research is real-time deployment. Even with a simple simulation, the model can be connected to live network traffic systems to evaluate the performance of the model in the real world.

Finally, there is room for improvement in the hyper-parameter tuning. Changing factors like learning rate, batch size, and number of layers can lead to better performance and more stable training.

In summary, these improvements could lead to a higher accuracy, reliability and applicability of the intrusion detection system for operational uses.

Based on our current results, we believe that the performance of the model could be further improved by improving the class balance.

\FloatBarrier

\section{Conclusion}

We present an improved CNN-LSTM based intrusion detection system and evaluate its performance with network traffic data in this paper. The results show better detection performance and very good accuracy. The model has great potential for practical applications.

Compared to the original CNN-based approach in \cite{healthcare_cnn}, The proposed paradigm shows better flexibility and performance under challenging IoT network conditions.

The proposed model extends the CNN-based approach presented in \cite{healthcare_cnn} It uses an LSTM layer to learn temporal features and perform multi-class classification, which improves the detection performance.

\end{document}